\documentclass[12pt,preprint]{aastex}

\shorttitle{The Fermi Bubbles}
\shortauthors{Fujita et al.}

\begin{document}


\title{A Hadronic-Leptonic Model for the Fermi Bubbles: Cosmic-Rays in
the Galactic Halo and Radio Emission}

\author{Yutaka Fujita}
\affil{Department of Earth and Space Science, Graduate School of
 Science, Osaka University, Toyonaka, Osaka 560-0043, Japan}
\email{fujita@vega.ess.sci.osaka-u.ac.jp}

\and

\author{Yutaka Ohira and Ryo Yamazaki}
\affil{Department of
Physics and Mathematics, Aoyama Gakuin University, Fuchinobe, Chuou-ku,
Sagamihara 252-5258, Japan}

\begin{abstract}
We investigate non-thermal emission from the {\it Fermi} bubbles on a
 hadronic model. Cosmic-ray (CR) protons are accelerated at the forward
 shock of the bubbles. They interact with the background gas in the
 Galactic halo and create $\pi^0$-decay gamma-rays and secondary
 electrons through proton-proton interaction. We follow the evolution of
 the CR protons and electrons by calculating their distribution
 functions. We find that the spectrum and the intensity profile of
 $\pi^0$-decay gamma-rays are consistent with observations. We predict
 that the shock front is located far ahead of the gamma-ray boundary of
 the {\it Fermi} bubbles. This naturally explains the fact that a clear
 temperature jump of thermal gas was not discovered at the gamma-ray
 boundary in recent {\it Suzaku} observations. We also consider
 re-acceleration of the background CRs in the Galactic halo at the shock
 front.  We find that it can significantly affect the gamma-rays from
 the {\it Fermi} bubbles, unless the density of the background CRs is
 $\lesssim 10$\% of that in the Galactic disk. We indicate that
 secondary electrons alone cannot produce the observed radio emission
 from the {\it Fermi} bubbles. However, the radio emission from the
 outermost region of the bubbles can be explained, if electrons are
 directly accelerated at the shock front with an efficiency of $\sim
 0.1$\% of that of protons.

\end{abstract}


\section{Introduction}

The {\it Fermi} bubbles are huge gamma-ray structure detected by the
{\it Fermi} gamma-ray satellite \citep{dob10,su10a,su12b}. The two
bubbles are symmetric about the Galactic plane and they extend $\sim
50^\circ$ above and below the Galactic center (GC). Their surface
brightness is almost uniform and they have sharp rims. The gamma-ray
spectrum is hard ($dN/dE\sim E^{-2}$). Similar structures have been
found in the radio \citep{fin04a,dob12b,pla13b} and X-ray bands
\citep{sno95a,sof00a}. While the morphology of the bubbles suggests that
they were created through some violent activities around the GC, the
origin of the cosmic-rays (CRs) that are responsible for the non-thermal
emission has not been identified.

Several models have been proposed to explain properties of the {\it
Fermi} bubbles. In some models, star formation activities around the GC
are thought to be the origin of the CRs \citep{cro11a,cro12a}. In other
models, CRs are carried by jets or outflows generated by the
super-massive black hole at the GC \citep{zub11a,guo12a,yan12a} or they
are accelerated in the bubbles \citep{mer11a,che11a}. Alternatively, the
{\it Fermi} bubbles may be a result of diffusive injection of Galactic
CR protons during their propagation through the Galaxy, if the bubbles
are expanding very slowly \citep{tho13a}. The gamma-rays can be
generated by interaction between CR protons and ambient gas (hadronic
models), or by inverse Compton scattering by CR electrons (leptonic
models).

\citet[][hereafter Paper~I]{fuj13d} proposed that CR protons accelerated
at the forward shock of the bubbles generate gamma-rays through the
hadronic interaction with gas protons ($pp$-interaction). We showed that
the observed gamma-ray properties of the {\it Fermi} bubbles, such as
the flat intensity profile, the sharp edge, and the hard spectrum, can
be reproduced if the CRs were accelerated when the bubbles were small,
and if the time scale of the energy injection at the GC was much smaller
than the current age of the bubbles. The CRs are confined in the
bubbles, because they excite Alfv\'en waves around the bubbles through
streaming instabilities and the waves scatter the CRs well.

In this paper, we further develop the hadronic model. We study
re-acceleration of background CRs by the {\it Fermi} bubbles. Those
background CRs have escaped from the Galactic disk, and they are
distributed in the Galactic halo. If the density of the background CRs
is large and/or if they are effectively re-accelerated, the gamma-ray
emission from the bubbles would be significantly affected by them. We
also investigate radio emission from CR electrons in the {\it Fermi}
bubbles, which was not investigated in Paper~I. This paper is organized
as follows. In Section \ref{sec:model}, we describe our model for the
evolution of the bubble and the acceleration of the CRs. In
Section~\ref{sec:result}, we show the results of our calculations and
discuss the re-acceleration of the background CRs and the radio emission
from the bubbles. Finally, Section~\ref{sec:conc} is devoted to
conclusions.

\section{Models}
\label{sec:model}

\subsection{Hydrodynamics of the Halo Gas}
\label{sec:hydro}

After energy is injected at the GC, a forward shock propagates in the
Galactic halo. We do not consider the details of the energy injection;
it may be caused by the activities of the super-massive black hole at
the GC. In Paper~I, we approximately treated the shock propagation using
self-similar solutions under the assumption that the shock is strong. In
this study, we numerically calculate the evolution, which enables us to
include the effects of gravity from the Galaxy and decreasing Mach
number of the shock. We assume that the time-scale of the energy
injection is much smaller than the current age of the bubbles ($\sim
10^7$~yr), because long-time injection is inconsistent with the
gamma-ray intensity profile (Paper~I). We do not consider the feedback
from the accelerated CRs on gas because the Mach number is small
($\lesssim 10$; see Figure~\ref{fig:mach}).

For the sake of simplicity, we assume that the bubble is spherically
symmetric, and we mainly focus on the high-galactic-latitude part of the
{\it Fermi} bubbles (large $|b|$ and small $|l|$ in the Galactic
coordinate). We solve the following equations:
\begin{equation}
\label{eq:cont}
 \frac{\partial \rho}{\partial t} +
  \frac{1}{r^2}\frac{\partial}{\partial r}(r^2\rho u)=0\:,
\end{equation}
\begin{equation}
 \frac{\partial(\rho u)}{\partial t} +
  \frac{1}{r^2}\frac{\partial}{\partial r}(r^2\rho u^2)
= -\rho\frac{\partial \Phi}{\partial r} 
- \frac{\partial P}{\partial r}\:,
\end{equation}
\begin{equation}
\label{eq:ene}
 \frac{\partial e}{\partial t} + \frac{1}{r^2}\frac{\partial}{\partial
r}[r^2 u (P + e)] = - \rho u \frac{\partial \Phi}{\partial r}\:,
\end{equation}
where $\rho$ is the gas density, $u$ is the velocity, $P$ is the
pressure, and $\Phi$ the Galactic potential. The total energy is defined
as $e=P/(\gamma-1)+\rho u^2/2$, where $\gamma=5/3$ is the adiabatic
index for the gas.

We use a fixed Galactic potential adopted by \citet{yan12a}. The
potential along the rotation axis of the Galaxy can be written as
\begin{equation}
 \Phi = \Phi_{\rm halo} + \Phi_{\rm disk} + \Phi_{\rm bulge}\:,
\end{equation}
where
\begin{equation}
 \Phi_{\rm halo}(r) = v_{\rm halo}^2\ln(r^2 + d_h^2)
\end{equation}
is the halo potential,
\begin{equation}
 \Phi_{\rm disk}(r) = -\frac{G M_{\rm disk}}{a+\sqrt{r^2 + b^2}}
\end{equation}
is the potential for the Miyamoto-Nagai disk \citep{miy75a}, and
\begin{equation}
 \Phi_{\rm bulge}(r) = -\frac{G M_{\rm bulge}}{r+d_b}
\end{equation}
is the potential for the Hernquist stellar bulge. We assume that $v_{\rm
halo}=131.5\:\rm km\: s^{-1}$, $d_h=12$~kpc, $M_{\rm disk}=10^{11}\:
M_\sun$, $a=6.5$~kpc, $b=0.26$~kpc, $M_{\rm bulge}=3.4\times 10^{10}\:
M_\sun$, and $d_b=0.7$~kpc. At $t=0$, the gas in the potential is
isothermal and in hydrostatic equilibrium. The temperature is
$T_h=2.4\times 10^6$~K and the density at $r=0$ is $\rho_0=1.0\times
10^{-23}\rm\: g\: cm^{-3}$.

\subsection{Cosmic-Rays}
\label{sec:CR}

In Paper~I, we solved a diffusion-convection equation to follow the
evolution of CR distribution functions $f_i(r,p,t)$, where $p$ is the
momentum of CRs. The equation was applied in the region far enough
away from the shock and the CR acceleration at the shock front was
separately treated by adopting a simple model. In this study, we treat
not only protons ($i=0$) but also electrons ($i=1$) and positrons
($i=2$). The equations are
\begin{equation}
\label{eq:diff}
 \frac{\partial f_i}{\partial t} 
= \frac{1}{r^2}\frac{\partial}{\partial r}\left(r^2 \kappa\frac{\partial
					       f_i}{\partial r}\right)
- (u+u_w)\frac{\partial f_i}{\partial r}
- \frac{1}{p^2}\frac{\partial}{\partial p}(p^2 \dot{p} f_i)
+ \frac{1}{3 r^2}\left[\frac{\partial}{\partial r}(r^2 (u+u_w))\right]
p\frac{\partial
f_i}{\partial p} + Q_i\:,
\end{equation}
where $\kappa$ is the diffusion coefficient, and $u_w$ is the velocity
of the waves that scatter CRs. The source $Q_i$ describes particle
injection for $r\neq R_{\rm sh}$, where $R_{\rm sh}$ is the shock
radius. The third term in the right-hand side represents cooling. For
protons, the cooling can be ignored, and hence $\dot{p}=0$. Moreover, we
assume that $Q_0=0$ for protons, because they are injected (accelerated)
only at the shock front. We also assume that $u_w=v_A$ for $r>R_{\rm
sh}$, where $v_A$ is the Alfv\'en velocity, and that $u_w=0$ for
$r<R_{\rm sh}$ because the waves would isotropically propagate there.

In Paper~I, we considered the evolution of $\kappa$ by calculating
growth of Alfv\'en waves through streaming instabilities. The results
showed that CRs are well scattered by the waves and confined around the
bubbles because of decrease of $\kappa$. While this also happens around
supernova remnants \citep{fuj10a,fuj11a}, the inefficiency of the CR
diffusion is more significant for the {\it Fermi} bubbles because of
their huge size for a given diffusion scale. Thus, we assume that the
diffusion coefficient $\kappa$ is much smaller than the Galactic value
(say by an order of magnitude as is shown in Figure~3 in Paper~I). In
this case, the results are not much different from the one with
$\kappa\approx 0$. Therefore, we take $\kappa\approx 0$ from now
on. Note that $\kappa$ is the diffusion coefficient outside a very
narrow region around the shock where the coefficient could be close to
the Bohm limit and CR are accelerated to high energies. We do not
explicitly solve CR acceleration there, and $\kappa$ is generally much
larger than the Bohm limit.

CRs accelerated at the shock front are swept downward from the shock
with the gas. In Paper~I, we considered only protons accelerated at the
shock front ($r=R_{\rm sh}$). In this study, we also investigate
re-acceleration of CRs in the Galactic halo. Thus, instead of giving
$Q_i$ at $r=R_{\rm sh}$, we approximate distribution functions at the
shock front, $f_{{\rm sh},i}=f_i(R_{\rm sh},p,t)$, based on a
steady-state, test-particle solution about a plane shock. Assuming that
the CRs are confined around the bubble, we can integrate a
diffusion-convection equation between the far upstream region of the
shock and $r=R_{\rm sh}$. We obtain
\begin{equation}
\label{eq:fshock0}
 p\frac{\partial 
f_{{\rm sh},i}}{\partial p} = 3\sigma (f_{\rm sh}-f_{b,i})\;,
\end{equation}
where $f_{b,i}(p)$ is the distribution function of the pre-existing CRs,
and $\sigma=\rho(r=R_{\rm sh}-0)/\rho(r=R_{\rm sh}+0)$ is the
compression ratio of the shock \citep{bla78a,dru83a}. From this
equation, we obtain
\begin{equation}
\label{eq:fshock}
f_{{\rm sh},i}
\approx q p^{-q}\int_{p_{{\rm inj},i}}^p p'^{q-1} f_{b,i}(p')dp'
+ f_{{\rm inj},i}\left(\frac{p}{p_{{\rm inj},i}}\right)^{-q}\:,
\end{equation}
where $f_{{\rm inj},i}$ is the normalization and $p_{{\rm inj},i}$ is
the lowest momentum, above which CRs can cross the shock; particles are
not accelerated for $p<p_{{\rm inj},i}$ \citep{bla78a,dru83a}. The index
is given by
\begin{equation}
\label{eq:q}
q = \frac{3\sigma}{\sigma - 1} = \frac{(\gamma + 1){\cal M}^2}{(\gamma -
 1){\cal M}^2 + 2}\:,
\end{equation}
where $\cal M$ is the Mach number of the shock \citep{bla87a}. The first
term in the right-hand side of equation~(\ref{eq:fshock}) refers to the
re-acceleration of CRs advected into the shock from upstream region, and
the second term represents CRs accelerated from the thermal background
plasma. The index $q$ ($>0$) is the increasing function of time in our
calculations.

For the injection of CR protons at the shock front, we adopt a thermal
leakage model \citep{kan02a,kan11a}:
\begin{equation}
\label{eq:finj0}
 f_{{\rm inj},0} = \frac{n_d}{\pi^{1.5}}p_{\rm th,0}^{-3}
\exp(-\xi_{{\rm inj},0}^2)\:,
\end{equation}
where $n_d$ is the downstream proton number density, $p_{\rm
th,0}=\sqrt{2 m_p k_B T_d}$ is the thermal peak momentum of the
downstream gas with a temperature of $T_d$, $m_p$ is the proton mass,
and $k_B$ is the Boltzmann constant. Here, $\xi_{{\rm inj},0}$ and
$p_{{\rm inj},0}$ are defined as
\begin{equation}
\label{eq:Qinj}
 \xi_{{\rm inj},0}\equiv \frac{p_{{\rm inj},0}}{p_{\rm th,0}}
\approx 1.17\frac{m_p u_d}{p_{\rm th,0}}
\left(1+\frac{1.07}{\epsilon_B}\right)
\left(\frac{\cal M}{3}\right)^{0.3}\:,
\end{equation}
where $u_d$ is the downstream flow speed in the shock rest frame. The
factor $\epsilon_B$ depends on the structure of magnetic fields and we
take $\epsilon_B=0.25$.

The acceleration of the CR protons is limited by the age of the
bubble. Since the acceleration time must be smaller than the age, the
maximum momentum is
\begin{equation}
\label{eq:p0max}
 p_{\rm max,0}=\frac{\sigma-1}{\sigma(\sigma+1)}
\frac{e B_b}{\eta_g c^2}V_{\rm
  sh}^2 t_{\rm age}\:,
\end{equation}
where $e$ is the proton charge, $B_b$ is the background magnetic field,
$\eta_g$ is the gyro factor, $V_{\rm sh}$ is the shock velocity, and
$t_{\rm age}$ is the bubble age \citep{aha99a,ohi10a}. Thus, for
$p>p_{\rm max,0}$, equation~(\ref{eq:fshock}) cannot be applied and
pre-existing CRs are just advected downstream. The maximum momentum
decreases as the bubble expands in our calculations.

Although electrons (primary electrons) should be accelerated at the
shock front in the same way as protons, their injection is expected to
be less efficient than that of protons. Thus, we assume that $f_{\rm
inj,1}=K_{\rm ep} f_{\rm inj,0}$, where $0\leq K_{\rm ep} < 1$ is the
parameter. The lowest energy of accelerated electrons, $p_{\rm inj,1}$,
is the same as $p_{\rm inj,0}$ but $m_p$ in equation~(\ref{eq:Qinj}) is
replaced by the electron mass $m_e$. The maximum momentum of electrons
is limited by cooling, mainly synchrotron emission and inverse Compton
(IC) scattering. The maximum momentum is obtained by replacing $t_{\rm
age}$ in equation~(\ref{eq:p0max}) by $t_{\rm cool}=|p/\dot{p}|$ :
\begin{equation}
\label{eq:p1max}
 p_{\rm max,1}=\frac{\sigma-1}{\sigma(\sigma+1)}
\frac{e B_b}{\eta_g c^2}V_{\rm
  sh}^2 t_{\rm cool}\:.
\end{equation}
We ignore positrons in the thermal plasma, and thus $f_{\rm
inj,2}=0$. The minimum and maximum momentums for positrons are the same
as those for electrons ($p_{\rm inj,2}=p_{\rm inj,1}$ and $p_{\rm
max,2}=p_{\rm max,1}$). 

The CR proton spectrum in the Galactic disk is approximated by
\begin{equation}
\label{eq:Jd}
 J_{d,0}(E) = 2.2\left(\frac{E}{\rm GeV}\right)^{-2.75}\rm\:
  cm^{-2}\: s^{-1}\: GeV^{-1}\: sr^{-1}\:,
\end{equation}
for $E\geq 9$~GeV and $J_{d,0}(E)=J_{d,0}(9\:\rm GeV)$ for $E<9$~GeV,
where $E$ is the energy \citep[e.g.][]{gab09a,ner12a}. The CR proton
spectrum in the Galactic halo is given by $J_{h,0}=K_{\rm halo}
J_{d,0}$, where $K_{\rm halo}$ is the parameter ($0\leq K_{\rm
halo}<1$). We assume that $J_{h,1}=J_{h,2}=0$, because cooling times of
CR electrons and positrons are much shorter than the Galactic age ($\sim
10$~Gyr, see Figure~28 in \citealt{su10a}) and because they are not
supplied in the halo. The background distribution functions are written
as
\begin{equation}
 4\pi p^2 f_{b,i} = \frac{4\pi}{c}J_{h,i}\frac{dE}{dp}
=4\pi c J_{h,i}\frac{p}{E}\:.
\end{equation}

The production rates of secondary electrons and positrons ($Q_1$ and
$Q_2$, respectively) created through $pp$-interaction between CR protons
and gas protons are calculated using the code provided by \citet{kar08b}
and the gas density distribution obtained by hydrodynamic simulations
(Section~\ref{sec:hydro}). We do not consider re-acceleration of
secondary electrons at the shock front because most of them are
generated in the far downstream region of the shock and not in the
vicinity of the shock. In order to calculate the cooling ($\dot{p}$) and
radiative processes for electrons and positrons, we adopt the models of
\citet{fan08a}. We include synchrotron radiation, IC scattering, Coulomb
interaction, and Bremsstrahlung. For IC scattering, we approximate the
interstellar radiation field (ISRF) by four blackbody components at 2.7,
35, 3000, and 7000~K. On the Galactic rotation axis, their energy
densities at 5~kpc away from the GC are $2.5\times 10^{-7}$, $1.5\times
10^{-7}$, $3.2\times 10^{-7}$, and $3.0\times 10^{-7}\rm\: MeV\:
cm^{-3}$, respectively \citep{por05a,por08a}. The ISRF data in GALPROP
code\footnote{http://galprop.stanford.edu/} show that the energy density
of the ISRF except for the cosmic microwave background component (2.7~K)
scales as
\begin{equation}
 U_{\rm ISRF}(r) \propto \frac{1}{(1+(r/0.57{\:\rm kpc})^2)^{0.61}}
\end{equation}
along the rotation axis of the Galaxy, where $r$ is distance from the
GC. For the background magnetic fields outside of the bubble ($r>R_{\rm
sh}$), we adopt a standard model in GALPROP with a lower limit:
\begin{equation}
 B_b(r)=\max[B_{b0}\exp(-r/r_{b0}), {\rm 1\:n G}]\:,
\end{equation}
where $B_{b0}=30\rm\: \mu G$, and $r_{b0}=2$~kpc. For $r<R_{\rm
sh}$, we assume that $B_b=10\rm\: \mu G$, because turbulence inside the
bubble would increase magnetic fields. However, it is unlikely that the
magnetic pressure $P_B$ exceeds the thermal pressure $P$. In our
calculation shown later in Figure~\ref{fig:rho}, they are comparable at
$t=10$~Myr, and $P_B<P$ at $t<10$~Myr. For $B_b=10\rm\: \mu G$, the
cooling time of CR electrons responsible for synchrotron radiation
($\sim$~GHz) is comparable to the age of the bubble.

\section{Results}
\label{sec:result}

We solve equation~(\ref{eq:diff}) with a boundary condition $f_i=f_{{\rm
sh},i}$ at $r=R_{\rm sh}$ (equation~(\ref{eq:fshock})). The gradient of
$f_i$ is zero at the inner boundary ($r=0.1$~kpc). The density $\rho$
and velocity $u$ of the background gas are obtained by solving
equations~(\ref{eq:cont})--(\ref{eq:ene}). We take the current time at
$t_{\rm obs}=10$~Myr.

\subsection{Fiducial Model}

In this subsection, we show the results for the fiducial model
(Model~FD) that gives almost the same results as the fiducial model in
Paper~I. We put a kinetic energy of $E_{\rm GC}=2\times 10^{57}$~erg at
$t=0$ and $r<0.3$~kpc. We do not include the background CRs escaped from
the Galactic disk ($K_{\rm halo}=0$). We consider only secondary
electrons (and positrons) that are created through $pp$-interactions,
and we do not treat primary electrons that are directly accelerated at
the shock front ($K_{\rm ep}=0$). From now on, the term 'electron'
includes 'positron' unless otherwise mentioned.

\begin{figure}
\epsscale{.60} \plotone{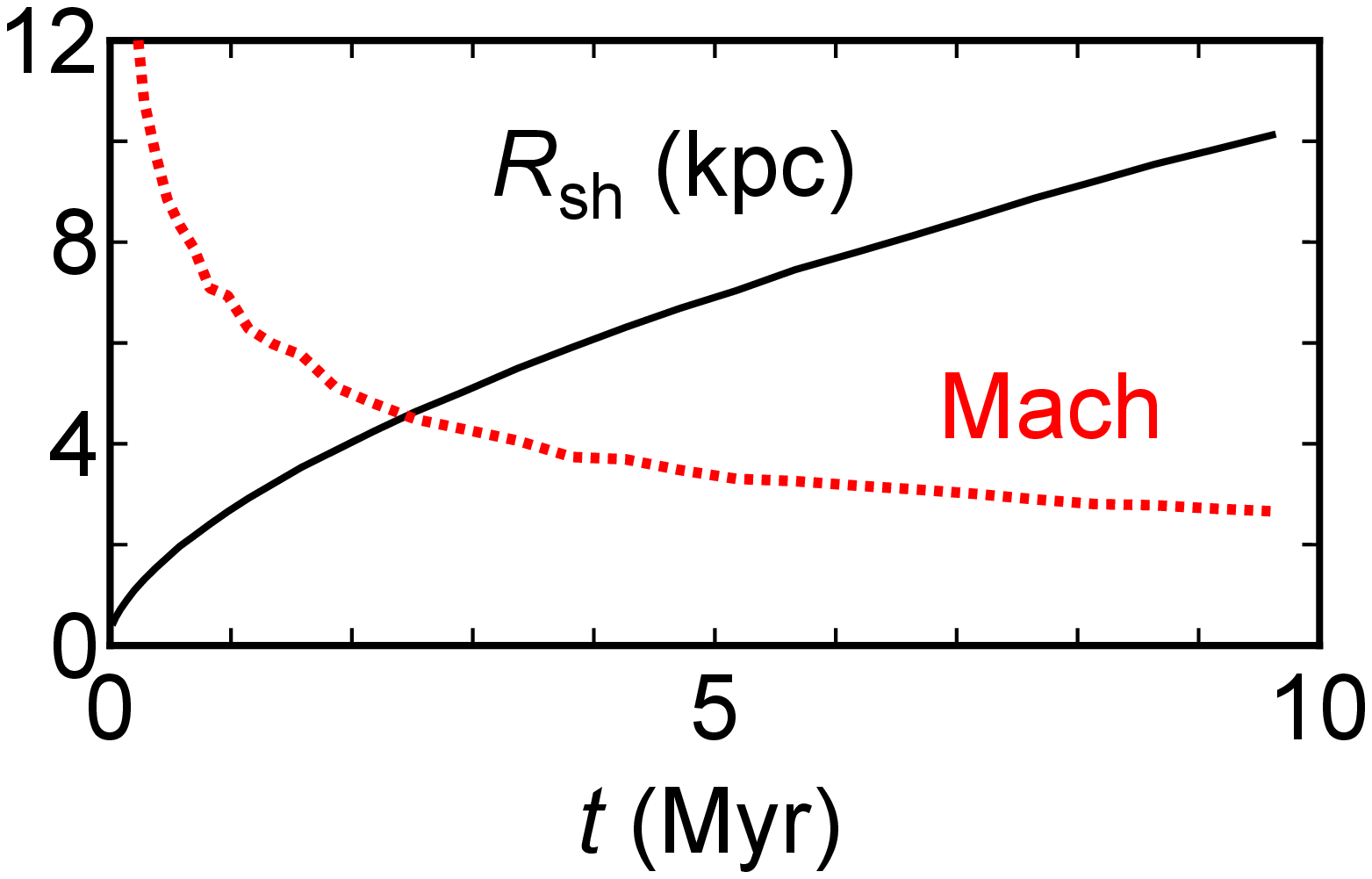} \caption{Evolution of the shock radius
 $R_{\rm sh}$ (solid) and the Mach number ${\cal M}$ (dotted) for Model
 FD.}\label{fig:mach}
\end{figure}

\begin{figure}
\epsscale{.60} \plotone{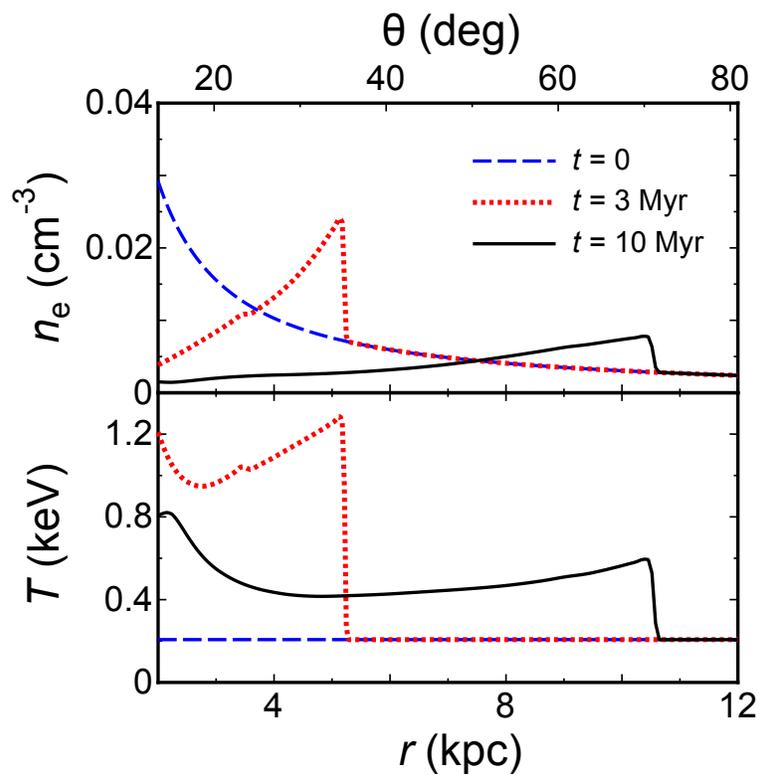} \caption{Thermal electron number density
 (upper) and temperature (lower) profiles of the background halo gas at
 $t=0$, 3 and 10~Myr for Model~FD.}\label{fig:rho}
\end{figure}

Figure~\ref{fig:mach} shows the evolution of the shock radius $R_{\rm
sh}$ and the Mach number $\cal M$. We start CR acceleration at
$t=0.5$~Myr. In Paper~I, we stopped the acceleration when ${\cal M}<4$,
because although a shock with a decreasing Mach number weakens the CR
acceleration, the model did not include that effect explicitly. In this
study, we include the effect by using
equation~(\ref{eq:finj0}). Therefore, less-efficient acceleration
continues even when ${\cal M}<4$. In Figure~\ref{fig:rho}, we show the
gas density (assumed to be the thermal electron number density)
and temperature profiles for Model~FD. The apparent size of the
bubble $\theta$ (degree) is simply given by $\theta = (180/\pi)r/d_{\rm
GC}$, where $d_{\rm GC}=8.5$~kpc is the distance to the GC.  As the Mach
number of the shock decreases, the density and temperature jumps at the
shock front also decrease. This effect was not included when we
calculated the gamma-ray emission through $pp$ interaction in
Paper~I. Figure~\ref{fig:rho} shows that some amount of gas remains even
far behind the shock, because the initial gas profile (at $t=0$) is
centrally concentrated (Paper~I).

In Figure~\ref{fig:gamma_FD}, we present the gamma-ray spectrum at
$t=t_{\rm obs}$ for Model~FD. The gamma-ray emission is originated from
the $\pi^0$-decay process associated with the $pp$-interaction. We show
the results for $\eta_g=1$ and 2000, which correspond to a large and a
small $p_{\rm max,0}$, respectively (equation~(\ref{eq:p0max})). Since
the result for $\eta_g=2000$ is consistent with the observations, we
adopt that value hereafter unless otherwise mentioned. Note that $p_{\rm
max,0}$ is related to the small diffusion coefficient of CRs in the
vicinity of the shock where CRs are accelerated. The coefficient there
is $\eta_g$ times the Bohm limit and is different from $\kappa$ in
equation~(\ref{eq:diff}). Although the value of $\eta_g=2000$ may
be rather large compared with $\eta_g\sim 1$ for supernova remnants in
the Galactic plane, we do not think that it is rejected for the {\it
Fermi} bubbles. We expect that seed magnetic fluctuations in the
Galactic halo are smaller than those in the Galactic plane because of
fewer sources of turbulence in the halo. This may prevent rapid growth
of the fluctuations around the shock of the {\it Fermi} bubbles and may
result in a larger $\eta_g$. Moreover, the Mach number of the shock of
the {\it Fermi} bubble (Figure~\ref{fig:mach}) is much smaller than that
of supernova remnants in the Galactic plane ($\sim$100). This would also
lead to the increase of $\eta_g$. The fairly hard spectra in
Figure~\ref{fig:gamma_FD} are mostly created by the CRs that are
accelerated when ${\cal M}$ is large, because the acceleration
efficiency decreases as ${\cal M}$ decreases. Since the CRs accelerated
at a shock with a large $\cal M$ have a hard spectrum with a small $q$
(equations~(\ref{eq:fshock}) and~(\ref{eq:q})), the gamma-ray spectrum
is also hard. The low energy part of the model spectrum ($\lesssim
6$~GeV) does not fit the date well (Figure~\ref{fig:gamma_FD}). This may
be because our model overestimates low energy CRs. Alternatively,
uncertainty of the data may be the reason as is shown by the discrepancy
between the data by \citet{su10a} and those by \citet{fra13}.

In Figure~\ref{fig:sp_FD}, we present the broad-band spectrum of
non-thermal emission from the bubble at $t=t_{\rm obs}$. The
$\pi^0$-decay gamma-rays come from the protons, while synchrotron, IC
scattering, and non-thermal Bremsstrahlung emissions come from the
secondary electrons. While the predicted gamma-ray spectrum is generally
consistent with the observations, the radio flux is smaller than that of
the radio observations. The deficiency of the radio luminosity to the
gamma-ray luminosity remains even if we change the background magnetic
fields, because the luminosity ratio intrinsically depends on the ratio
of the production rate of charged pions to that of neutral pions in the
$pp$-interaction \citep[e.g.][]{cro11a}. For the given magnetic fields
inside the bubble ($B_b=10\rm\:\mu G$), the cooling time of electrons is
comparable with the age of the bubble (Section~\ref{sec:CR}). Thus,
smaller $B_b$ gives smaller synchrotron luminosity because the
emissivity decreases. Larger $B_b$ does not lead to larger synchrotron
luminosity because radiative cooling reduces the number of high-energy
electrons.

Figure~\ref{fig:pro_FD} shows that the gamma-ray intensity profile is
consistent with the observed profile. The former is calculated simply by
projecting the gamma-ray emission on a plane at a distance of $d_{\rm
GC}=8.5$~kpc and we do not consider detailed geometrical effects that
come up when the distance to the {\it Fermi} bubbles is finite. 
This is because we assumed the spherical symmetry and the size of the
bubble is comparable to $d_{\rm GC}$. Thus, the detailed comparison with
the observations may be premature. However, the line of sight cross
section near the bubble edge, or the curvature of the bubble, does not
much differ between a more realistic model considering the position of
the Sun (e.g. Figure~4 of \citealt{sof00a}) and our model for a given
bubble volume. Since we do not include background gamma-rays, we shift
the observational data along the vertical axis ($-0.9\rm\: keV \:
cm^{-2}\: s^{-1}\: sr^{-1}$ in the 1--5~GeV band and $-0.4\rm\: keV \:
cm^{-2}\: s^{-1}\: sr^{-1}$ in the 5--20~GeV band). The brightness
profile is fairly flat for $\theta\lesssim 50^\circ$, because the bubble
is not empty of gas and the CR protons interact with the gas protons
contained in the bubble (Figure~\ref{fig:rho}, see also Paper~I). The
brightness rapidly decreases outside of the edge at $\theta\sim
50^\circ$, because the CRs confined in the bubble and do not diffuse out
(see Paper~I). Another reason is that the gas density is high just
behind the shock front (Figure~\ref{fig:rho}). The gamma-ray edge is
inside the shock front at $\theta\sim 70^\circ$
(Figure~\ref{fig:rho}). Recent X-ray observations with {\it Suzaku} have
shown that there is no clear temperature jump at the gamma-ray edge
\citep{kat13a}. This is consistent with this model prediction because
the edge does not reflect the shock front.  In Figure~\ref{fig:pro_FD},
the radio intensity profile is flat and it is smaller than the gamma-ray
one. Recent analyses of {\it WMAP} and {\it Planck} data have revealed
that the radio intensity profile is almost identical to the gamma-ray
profile around the edge of the {\it Fermi} bubbles (Figure~3 in
\citealt{dob12b} and Figure~10 in \citealt{pla13b}). This means that the
predicted radio surface brightness falls short of the observed one.

\subsection{Re-Acceleration of the Galactic Halo CRs}

In this subsection, we investigate re-acceleration of the CRs in the
Galactic halo.  Since our model is rather simple and we know little
about the amount of CRs and their distribution in the Galactic halo, we
only focus on whether a significant amount of CRs ($K_{\rm halo}\gtrsim
0.5$) affect the results or not. We assume that the background CRs are
uniform for simplicity, because equation~(\ref{eq:fshock0}) assumes that
the upstream CRs are uniformly distributed.

\begin{figure}
\epsscale{.60} \plotone{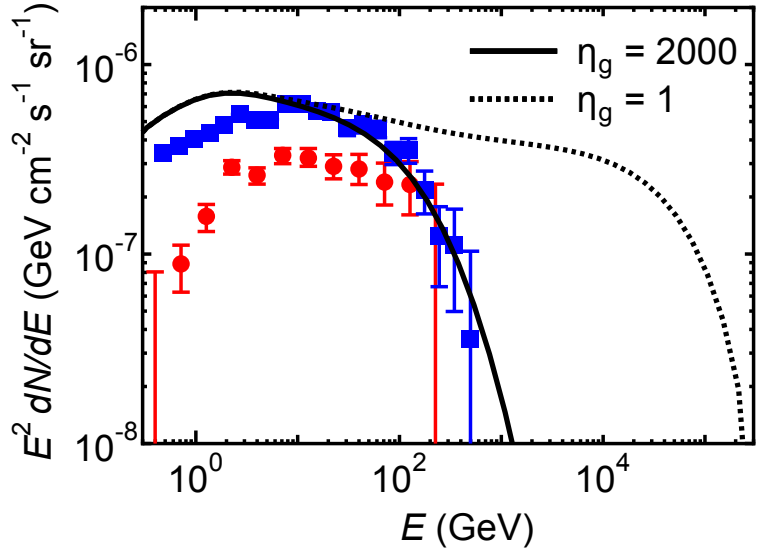} \caption{Gamma-ray spectrum at $t=t_{\rm
 obs}$ for Model~FD. The solid curve is for $\eta_g=2000$ and the dotted
 curve is for $\eta_g=1$. Filled circles are the {\it Fermi}
 observations by \citet{su12b}. Filled squares are preliminary results
 obtained by \citet{fra13}.}  \label{fig:gamma_FD}
\end{figure}

\begin{figure}
\epsscale{.60} \plotone{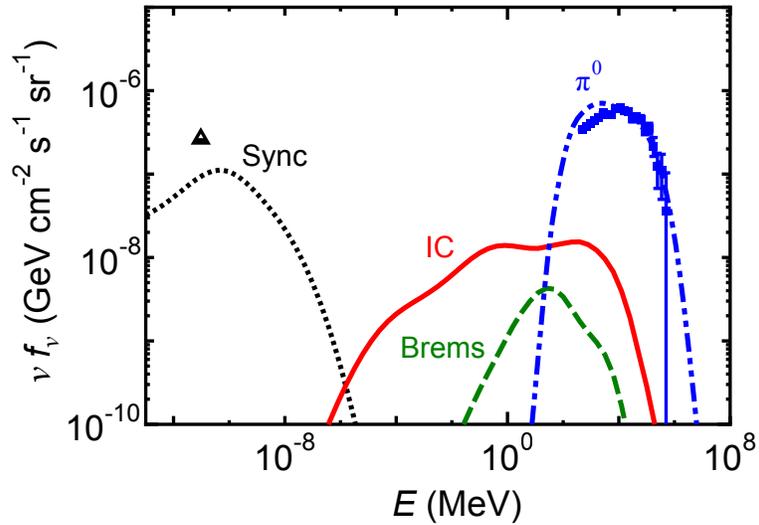} \caption{Broad-band spectrum at
 $t=t_{\rm obs}$ for Model~FD. Synchrotron radiation (dotted line), IC
 scattering (solid line), and non-thermal bremsstrahlung (dashed line)
 are of the electrons. $\pi^0$-decay gamma-rays are shown by the
 double-dot-dashed line. Filled squares are preliminary {\it Fermi}
 results obtained by \citet{fra13}. Open triangle is the observation by
 \citet{car13a}} \label{fig:sp_FD}
\end{figure}

\begin{figure}
\epsscale{.60} \plotone{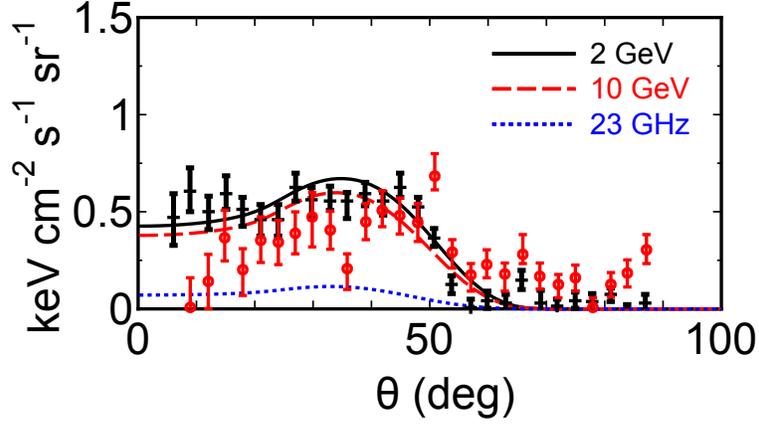} \caption{Gamma-ray (solid: 2~GeV, and
 dashed: 10~GeV) and radio (dotted: 23~GHz) intensity profiles at
 $t=t_{\rm obs}$ for Model~FD. The crosses (1--5 GeV band) and the
 circles (5--20 GeV band) are the observations for the southern bubble
 shown in Figure 9 of \citet{su10a}. Observed radio profile is almost
 identical to the gamma-ray profile around the edge of the {\it Fermi}
 bubbles (see text).}  \label{fig:pro_FD}
\end{figure}

\begin{figure}
\epsscale{.60} \plotone{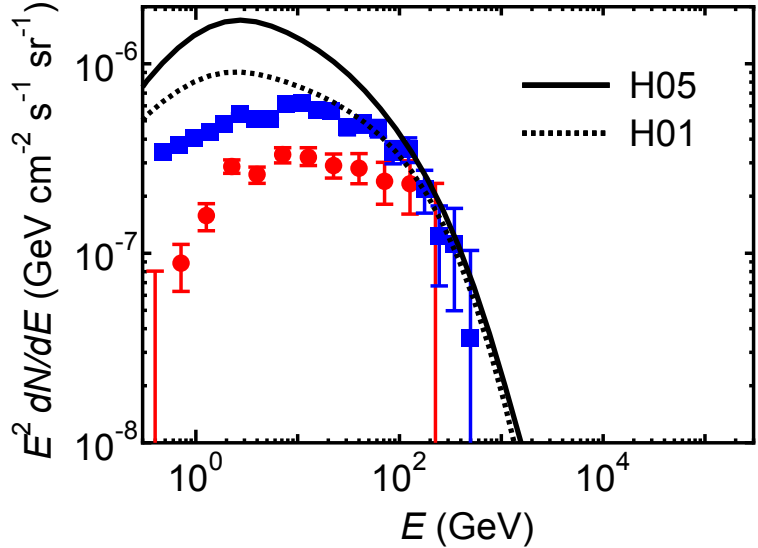} \caption{Same as
Figure~\ref{fig:gamma_FD} but for the re-acceleration models. The solid
curve is for Model~H05 and the dotted curve is for Model~H01.}
\label{fig:gamma_H}
\end{figure}

\begin{figure}
\epsscale{.60} \plotone{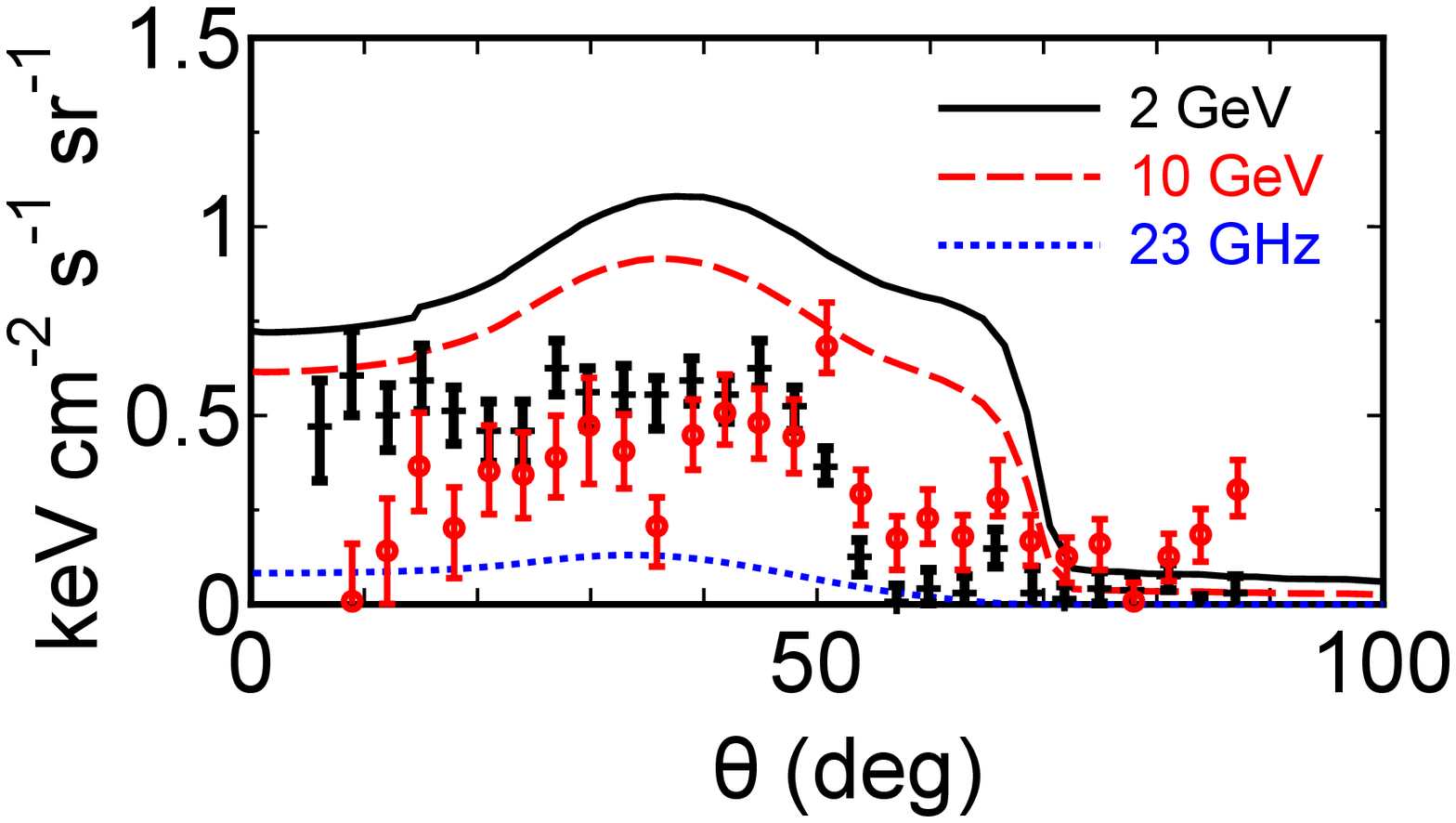} \caption{Same as 
Figure~\ref{fig:pro_FD} but for Model~H05.} \label{fig:pro_H05}
\end{figure}

Model~H05 is the same as Model~FD but for $K_{\rm halo}=0.5$.  In
Figure~\ref{fig:gamma_H}, we present the gamma-ray spectrum for
Model~H05. We ignore the emission from CRs at $r>R_{\rm sh}$. Compared
with Model~FD (Figure~\ref{fig:gamma_FD}), the gamma-ray flux at a few
GeV is larger in Model~H05. This is because the re-accelerated CRs also
contribute to the gamma-ray flux. In this model, the Galactic CRs
re-accelerated later with smaller $p_{\rm max,0}$ and larger $q$
contribute more to the gamma-rays than the CRs accelerated earlier from
the background thermal gas with larger $p_{\rm max,0}$ and smaller
$q$. Thus, for $3\lesssim E\lesssim 100$~GeV, the gamma-ray spectrum is
softer than that predicted in Model~FD and the observations
(Figures~\ref{fig:gamma_H} and~\ref{fig:gamma_FD}). The intensity
profile is displayed in Figure~\ref{fig:pro_H05}. The gamma-ray profile
has a shape edge at $\theta \sim 70^\circ$ (Figure~\ref{fig:pro_H05}),
which corresponds to the position of the shock front at $R_{\rm
sh}=10.4$~kpc (Figure~\ref{fig:rho}). This means that there should be a
temperature jump of a factor of three at the gamma-ray edge, which is
inconsistent with the {\it Suzaku} observations \citep{kat13a}. If we
take $K_{\rm halo}=0.1$ (Model~H01), the contribution from the
re-accelerated CRs becomes much smaller (Figures~\ref{fig:gamma_H}
and~\ref{fig:gamma_FD}). This means that the density of the background
CRs at the position of the {\it Fermi} bubbles ($\sim 5$--10~kpc above
the Galactic plane) must be much smaller than that in the Galactic disk.

\begin{figure}
\epsscale{.60} \plotone{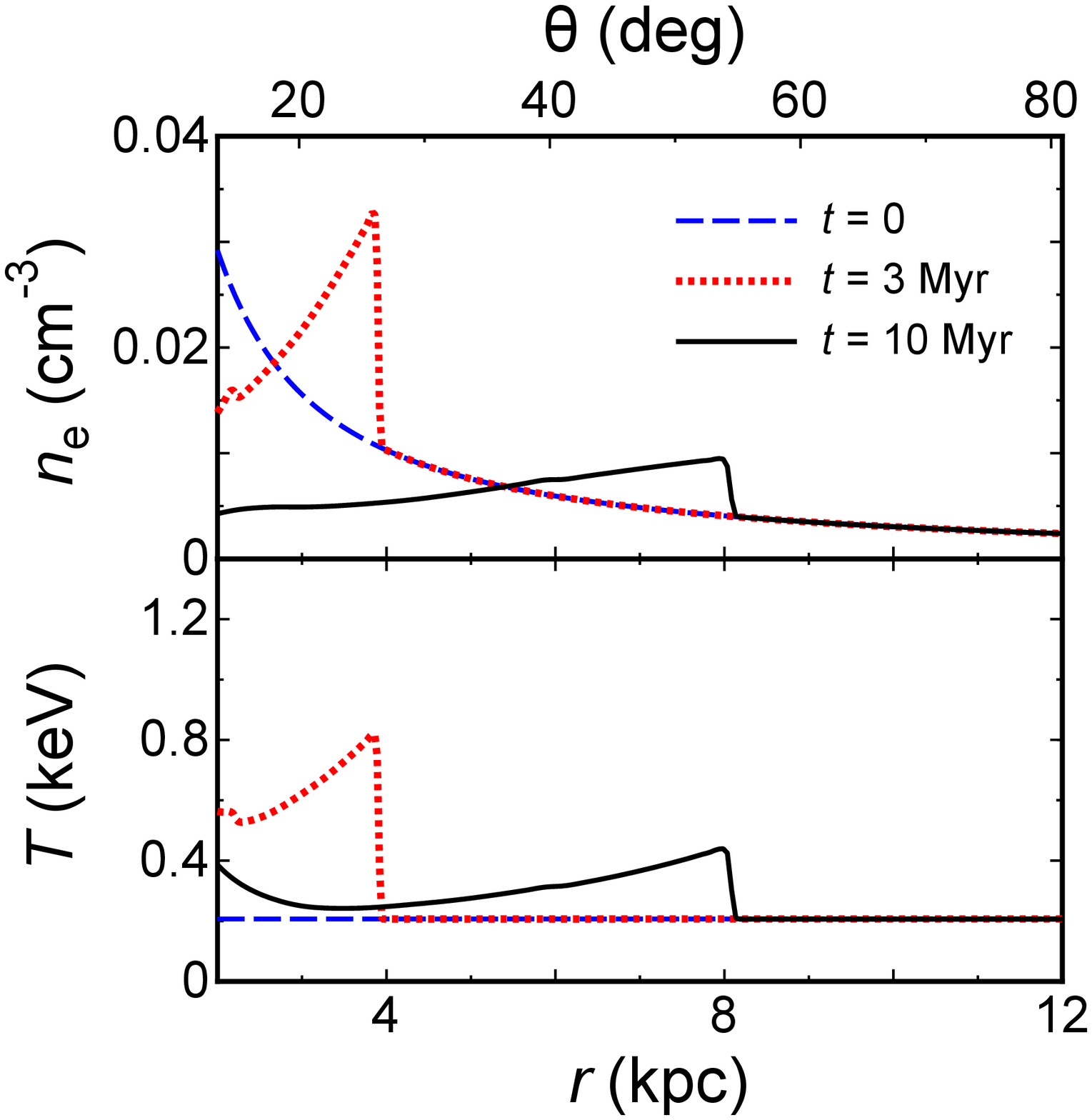} \caption{Same as Figure~\ref{fig:rho}
but for Model~HS.}\label{fig:rhoHS}
\end{figure}

\begin{figure}
\epsscale{.60} \plotone{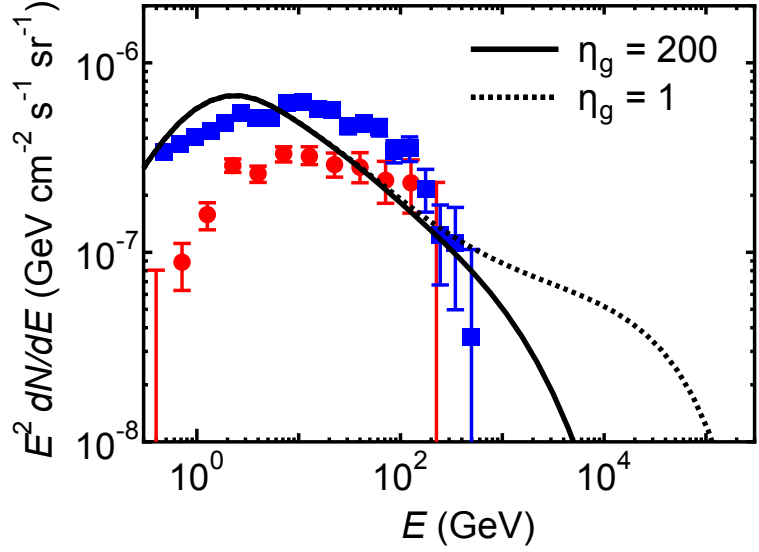} \caption{Same as
Figure~\ref{fig:gamma_FD} but for Model~HS. The solid curve is for
$\eta_g=200$ and the dotted curve is for $\eta_g=1$.}
\label{fig:gamma_HS}
\end{figure}

\begin{figure}
\epsscale{.60} \plotone{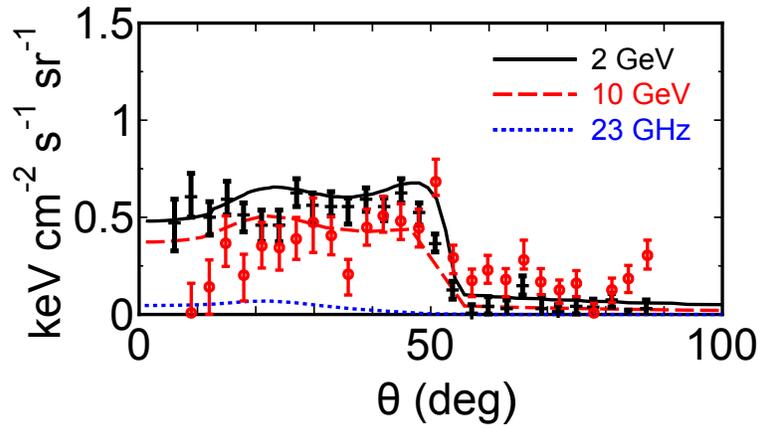} \caption{Same as 
Figure~\ref{fig:pro_FD} but for Model~HS ($\eta_g=200$).} 
\label{fig:pro_HS}
\end{figure}

We study another re-acceleration model in which the gamma-rays from the
{\it Fermi} bubbles are totally attributed to the re-accelerated CRs. In
this model (Model~HS), the input energy at the GC is chosen so that the
size of the gamma-ray bubble matches with the observed one. The CRs
accelerated from the thermal gas are not included ($f_{\rm inj,0}=f_{\rm
inj,1}=f_{\rm inj,2}=0$). We take $E_{\rm GC}=7\times 10^{56}$~erg and
$K_{\rm halo}=0.5$. Figure~\ref{fig:rhoHS} shows the density and
temperature profiles of the background halo gas. At $t=t_{\rm
obs}=10$~Myr, the shock front is at $R_{\rm sh}= 7.9$~kpc and the Mach
number is ${\cal M}=2.1$. In Figure~\ref{fig:gamma_HS}, we show the
gamma-ray spectra for $\eta_g=200$ and 1. As can be seen, they are not
much different in the GeV band. When $\eta_g=1$, the spectrum is softer
than that in Model~FD at $E\gtrsim 3$~GeV (Figures~\ref{fig:gamma_HS}
and~\ref{fig:gamma_FD}). In Model~FD, the gamma-rays are largely
originated from the CRs accelerated from the thermal gas when $t$ is
small or ${\cal M}$ is large. In Model~HS, on the contrary, the CRs
re-accelerated when $t$ is large or ${\cal M}$ is small contribute to
the gamma-rays. Thus, the gamma-ray spectrum is softer. Although the
gamma-ray profile reproduces the observations (Figure~\ref{fig:pro_HS}),
the predicted spectra in Figure~\ref{fig:gamma_HS} are softer than the
observations at $E\gtrsim 3$~GeV. In Model~HS, the shock front
corresponds to the gamma-ray edge (Figures~\ref{fig:rhoHS}
and~\ref{fig:pro_HS}). The temperature jump of a factor of two at the
shock front (Figure~\ref{fig:rhoHS}) is larger than that measured with
{\it Suzaku} at the gamma-ray edge \citep[a factor of $<
1.5$;][]{kat13a}. Even if we change $E_{\rm GC}$ and/or $t_{\rm obs}$,
we cannot adjust the gamma-ray spectrum and the gas temperature profile
with the observations at the same time.  Thus, we conclude that this
model cannot reproduce the observations. Moreover, the surface
brightness in the radio band is too small to be consistent with the
observations (Figure~\ref{fig:pro_HS}).

\subsection{Electron Acceleration and Radio Emission}
\label{sec:rad}

The models studied above cannot explain the observed radio emission. In
hadronic models, the radio luminosity is essentially smaller than the
gamma-ray luminosity \citep[e.g.][]{cro11a}. One idea to solve this
problem is to assume that CR electrons that are accelerated around the
GC are supplied to the inside of the {\it Fermi} bubbles. Alternatively,
the electrons may be accelerated within the bubbles. The synchrotron
radiation from the electrons may fill the gap in the radio luminosity
between the observations and the hadronic models.

\begin{figure}
\epsscale{.60} \plotone{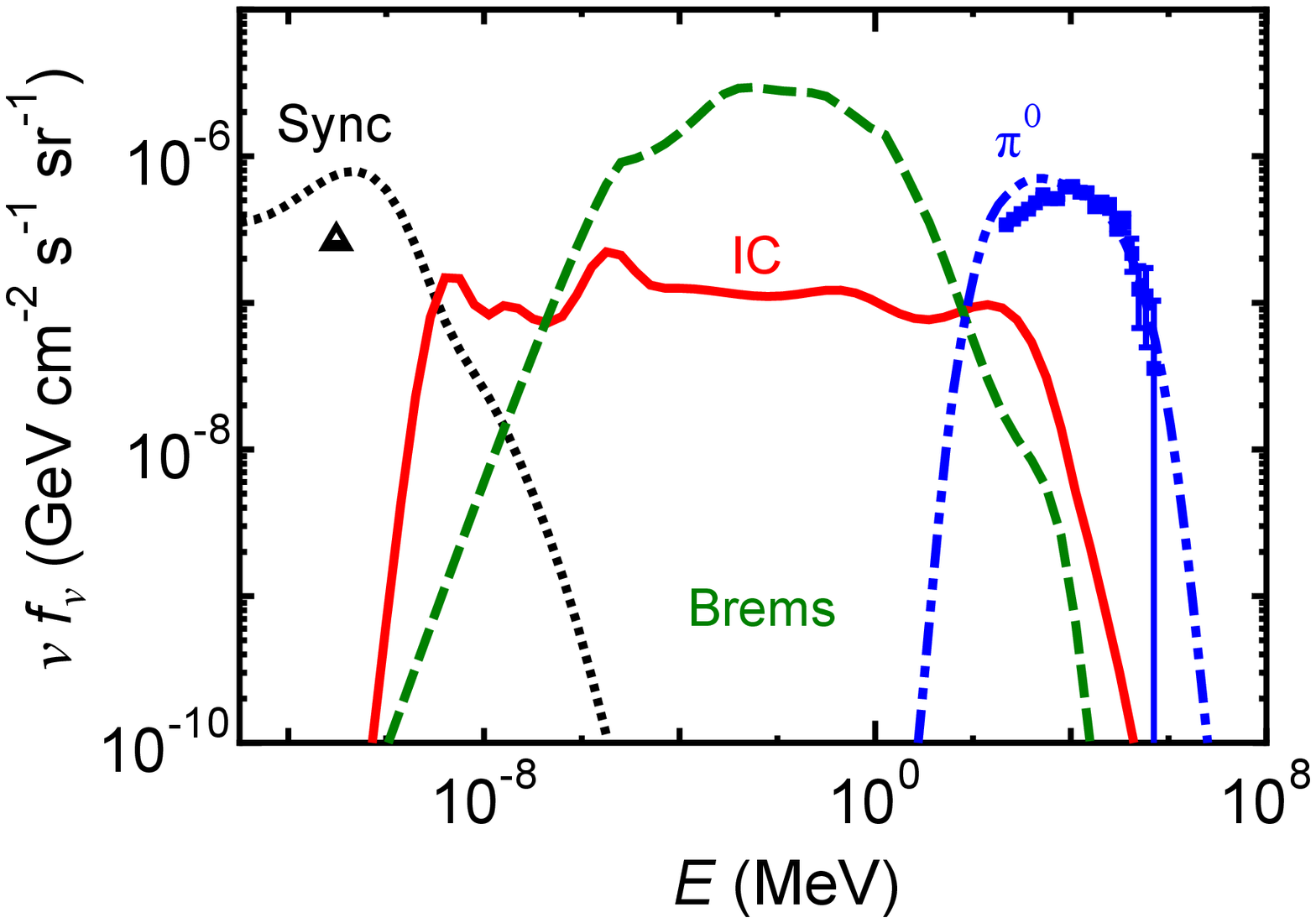} \caption{Same as 
Figure~\ref{fig:sp_FD} but for Model~EL.} \label{fig:sp_EL}
\end{figure}

\begin{figure}
\epsscale{.60} \plotone{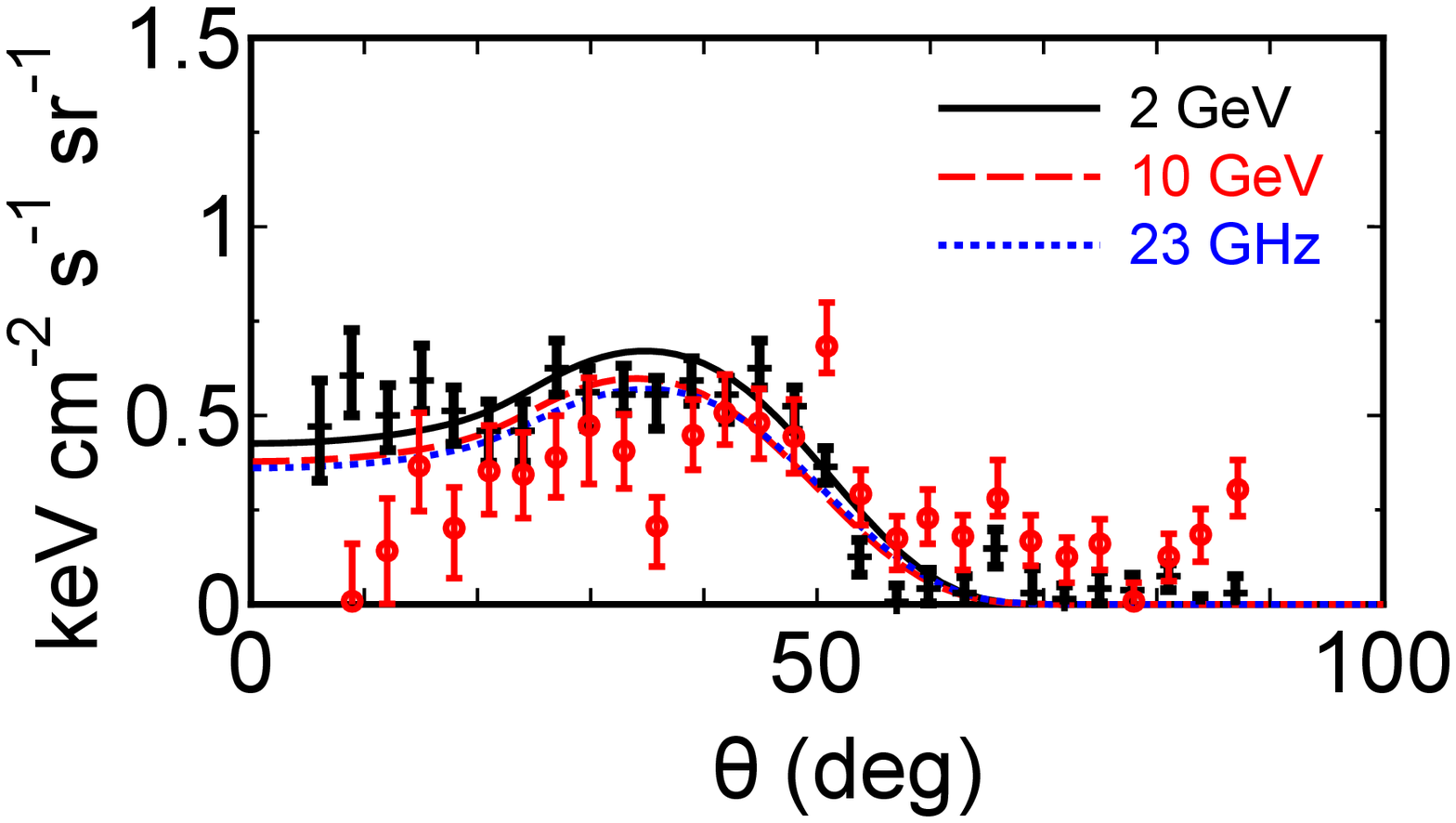} \caption{Same as 
Figure~\ref{fig:pro_FD} but for Model~EL.} \label{fig:pro_EL}
\end{figure}

Here, we consider another solution in which primary electrons are
included. Model~EL is the same as Model~FD but for $K_{\rm ep}=1.3\times
10^{-3}$. The value of $K_{\rm ep}$ is chosen to reproduce the
observations. The broad-band spectrum for Model~EL is shown in
Figure~\ref{fig:sp_EL}. The gamma-ray spectrum is identical to that for
Model~FD (Figure~\ref{fig:sp_FD}), because IC scattering hardly
contributes the emission in that band. On the contrary, the radio
luminosity is increased, owing to the synchrotron radiation from the
primary electrons. In Figure~\ref{fig:pro_EL}, we present the intensity
profiles in the gamma-ray and radio bands. Since the predicted radio
profile is almost the same as the gamma-ray profile around the bubble
edge, it is consistent with the observations \citep{dob12b,pla13b}.
Moreover, the spectral index at $\sim 30$--40~GHz is $-0.5$ (for $\nu
f_\nu$ in Figure~\ref{fig:sp_EL}). This is consistent with the {\it
Planck} observations at the bubble edge. Actually, Figure~10 of
\citet{pla13b} shows that the radio surface brightness at 43~GHz is a
factor of 1.15 smaller than that at 30~GHz.

However, the {\it Planck} observations have shown that for
$b>-35^{\circ}$, the radio surface brightness is significantly larger
than that for $b< -35^{\circ}$ (Figure~9 in \citealt{pla13b}). As is
mentioned in Section~\ref{sec:hydro}, we focus on the
high-galactic-latitude part of the {\it Fermi} bubbles, and the inner
emission cannot be reproduced by this model. It may be generated by
freshly injected electrons from the GC. The spectrum index of the whole
{\it Fermi} bubbles in the {\it Planck} band is 0.5 \citep[for $\nu
f_\nu$;][]{pla13b}. Most of the emission may be associated with the
inner bright emission.  Note that the radio luminosity in
Figure~\ref{fig:sp_EL} is a factor of a few larger than the observations
\citep{car13a}, while the radio profile is consistent with the radio
observations (Figure~\ref{fig:pro_EL}). This may be owing to the broader
region \citet{car13a} studied or the difference of the background
estimation.

\begin{figure}
\epsscale{.60} \plotone{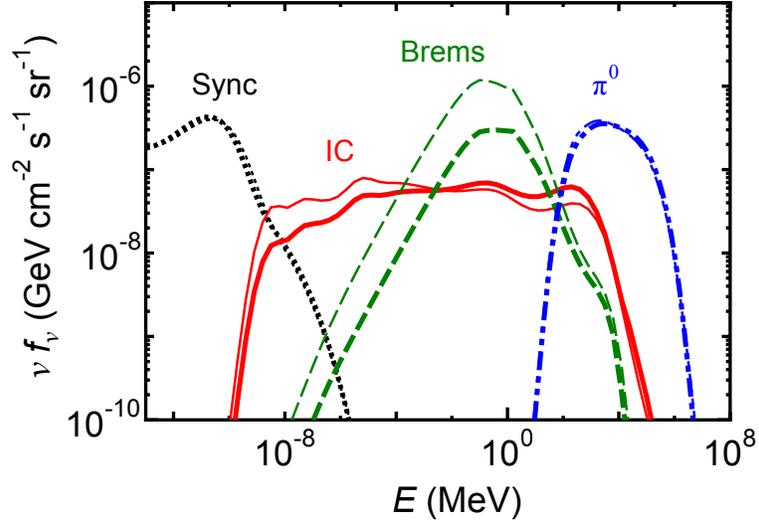} \caption{Broad-band spectra at
 $t=t_{\rm obs}$ for Model~EL. The thick lines are for $r<0.68\: R_{\rm
 sh}$ and the thin lines are for $0.68\: R_{\rm sh}<r<0.9\: R_{\rm
 sh}$. Synchrotron radiation (dotted line), IC scattering (solid line),
 and non-thermal bremsstrahlung (dashed line) are of the electrons.
 $\pi^0$-decay gamma-rays are shown by the double-dot-dashed line.}
 \label{fig:sp_EL2}
\end{figure}

\begin{figure}
\epsscale{.60} \plotone{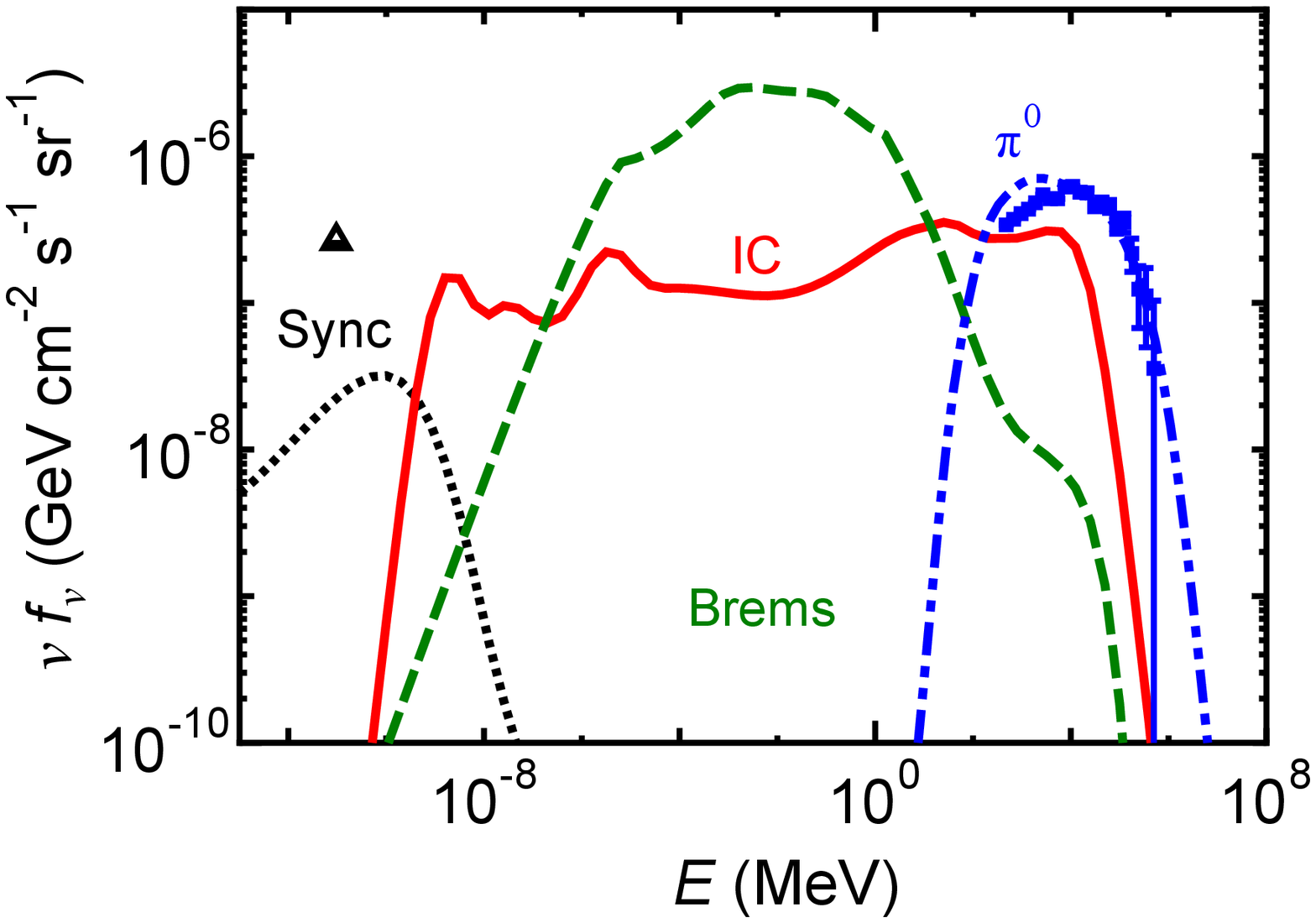} \caption{Same as
Figure~\ref{fig:pro_FD} but for Model~ELS.}  \label{fig:sp_ELS}
\end{figure}

Figure~\ref{fig:sp_EL2} shows the spectra for $r<0.68\: R_{\rm sh}$ and
for $0.68\: R_{\rm sh}<r<0.9\: R_{\rm sh}$. The effect of projection is
not included. Note that the shock radius corresponds to
$\theta=70^\circ$.  The spectra in the inner and the outer regions are
almost the same in the gamma-ray ($\pi^0$-decay) and radio (synchrotron)
bands. The IC scattering and Bremsstrahlung spectra are a little harder
in the inner region, because the Mach number of the shock is larger and
the spectrum of the accelerated electrons is harder when $R_{\rm sh}$ is
smaller. Since $q$ in equation~(\ref{eq:fshock}) is the increasing
function of time, we cannot ignore the contribution of lower-energy CR
electrons ($p\sim p_{\rm inj,1}$) that are recently accelerated at the
shock to the IC scattering and Bremsstrahlung emissions
(Figure~\ref{fig:pro_EL}). Figure~\ref{fig:sp_ELS} shows the
spectra of Model~ELS, which is the same as Model~EL but $B_b=1\rm\:\mu
G$. While the IC luminosity is larger than that in Model~EL, the
synchrotron luminosity is smaller and cannot reproduce radio
observations.

\section{Conclusions}
\label{sec:conc}

We have studied non-thermal radiation from the {\it Fermi} bubbles. In
our model, CR protons are accelerated at the forward shock of the
bubbles. They produce gamma-rays and secondary CR electrons via
$pp$-interaction. We followed the evolution of the distribution
functions of the CRs. Our fiducial model can reproduce the observed hard
spectrum and flat intensity profile in the gamma-ray band, because most
CRs are accelerated at early times when the Mach number of the shock is
large and because the bubble is not empty of gas. In this model, the
edge of the gamma-ray bubble does not correspond to the shock
front. This is consistent with recent {\it Suzaku} observations showing
that there is no temperature jump at the edge. However, the predicted
radio flux from secondary electrons is much smaller than the
observations.

The Galactic halo may be filled with background CRs escaped from the
Galactic disk. Thus, we investigated re-acceleration of those CRs by the
{\it Fermi} bubbles. We found that the gamma-rays from the
re-accelerated CRs significantly affect the luminosity and the intensity
profile, which may indicate that the density of the background CRs in
the halo is $\lesssim 10$\% of that in the disk. Although the observed
gamma-ray intensity profile can be reproduced by a model in which only
re-accelerated CRs are included, the model contradicts the observed hard
gamma-ray spectrum and the non-detection of X-ray temperature jump at
the gamma-ray edge.

Since secondary electrons alone cannot explain the observed radio
luminosity, we include primary electrons that are accelerated from
thermal plasma at the shock. We found that the radio intensity profile
around the bubble edge can be reproduced if electrons are accelerated
with an efficiency of $\sim 0.1$\% of that of protons.

\acknowledgments

We are grateful to the referee for valuable comments. We thank M.~Ozaki
for useful comments. This work was supported by KAKENHI (YF: 23540308,
YO: No.24.8344). R.~Y. was supported by the fund from Research
Institute, Aoyama Gakuin University.

\clearpage

\end{document}